\documentstyle[revtex,eqsecnum]{aps}

\begin{document}

\begin{title}
Relativistic three-particle scattering equations
\end{title}
\author{Sadhan K. Adhikari and Lauro Tomio}
\begin{instit}
Instituto de F\'\i sica Te\'orica, Universidade Estadual Paulista \\
01405-000 S\~{a}o Paulo, S\~{a}o Paulo, Brasil
\end{instit}
\moreauthors{T. Frederico}
\begin{instit}
Instituto de Estudos Avan\c cados, Centro T\'ecnico Aeroespacial \\
12231-970 S\~ao Jos\'e dos Campos,  S\~{a}o Paulo, Brasil
\end{instit}

\begin{abstract}
We derive a set of relativistic three-particle
scattering equations in the three-particle c.m. frame employing
a relativistic three-particle
propagator suggested long ago by Ahmadzadeh and Tjon in the c.m. frame of a
two-particle subsystem. We make the coordinate transformation of this
propagator from the c.m. frame of the two-particle subsystem to the
three-particle c.m. frame. We also point out that some  numerical
applications of the Ahmadzadeh and Tjon propagator to the
three-nucleon problem use unnecessary nonrelativistic approximations which
do not simplify the computational task, but violate constraints of
relativistic unitarity and/or covariance.

\end{abstract}
\vskip 0.5cm

There are a variety of approaches for writing three-particle relativistic
scattering equations.\cite{RT1,BLS,AAY,WALL,AFNAN,GLOK,FUDA}
About 25 years ago in a classic paper Ahmadzadeh and Tjon\cite{TJON1} (AT)
suggested a relativistic three-particle propagator in the
center of momentum (c.m.) frame of a two-particle subsystem of the three
particles [their Eq. (2.12)], using a dispersion relation
in the two-particle invariant energy, to be used in relativistic three-particle
scattering equations.  The AT propagator in the two-particle c.m. frame
satisfies constraints of relativistic covariance and unitarity.
Here we derive a relativistic
three-particle scattering equation  using the transformed
AT propagator in the three-particle c.m. frame.

Also, there have been several numerical calculations
on studies of the relativistic effect in the three-nucleon system using the AT
propagator.\cite{TJON2,RINAT,GARC,TJON3,MACH}  The original AT propagator was
derived in the c.m. frame of the two-particle subsystem. However,
these numerical  calculations employed relativistic three-particle
scattering equations in the three-particle c.m. frame. One needs to
transform the AT propagator to the three-particle c.m. frame for this
purpose. The three-particle relativistic propagators used in these numerical
calculations\cite{TJON2,RINAT,GARC,TJON3,MACH} involve
unwanted and unnecessary nonrelativistic approximations, which
do not facilitate the  computational task but    violate
constraints of relativistic unitarity and/or covariance.

Relativistic three-particle scattering equations used in these studies
are diagramatically represented in Fig. 1. Each single-line represents
propagation of a particle, and double-line represents propagation of
the bound state or isobar of two particles. The labels in the figure
represent four-momentum of the propagating particle; $P$ is the total
four-momentum of the system of three particles in the three-particle c.m.
frame and is given by $(\sqrt s,0,0,0)$. Usual
relativistic three-particle scattering equations employ a three-dimensional
reduction of this equation in the form
\begin{equation}
T(\vec q, \vec k, s)=2B(\vec q,\vec k,s)+\frac {2}{(2\pi )^3}\int
\frac{d\vec p}{2\omega _{\vec p}}
B(\vec q, \vec p,s)  \tau (\sigma _p) T(\vec p, \vec k, s).
\label{1K} \end{equation}
Here we consider three equal-mass particles of mass $m$, so that
$\omega_{\vec p}=(|\vec p|^2+m^2)^{1/2}$, etc. In Eq. (\ref{1K}) $T$ is the
$t$ matrix for scattering of a particle from the bound state or isobar of two
particles, $\tau$ is the dressed propagator for the two-particle bound state
or the isobar, and $g$'s are the vertex function for  these  bound states
or isobars.

The AT three-particle propagator is given, in the c.m. frame of the
two-particle subsystem, with two particles of
momentum $\vec p$ and $-(\vec p+\vec q)$ in the three-particle c.m. frame
(see, Fig. 1), respectively, by
[Eq. (2.12) of AT]:
\begin{equation}
G =  \frac  {\hat \omega_{\vec p}
+ \hat \omega_{\vec p+\vec q}} {\hat \omega_{\vec p}
 \hat \omega_{\vec p+\vec q}}
\frac {2} {(\sqrt s - \omega_{\vec q})^2 - {|\vec q|}^2 - (\hat \omega_{\vec p}
+ \hat \omega_{\vec p+\vec q})^2+i0}.
\label{2}
\end{equation}
Here we have changed the $i,j,k$ labels of AT to explicit momentum labels
for identifying particles. The hats over the $\omega$'s mean that the
corresponding energies are to be calculated in the two-particle c.m. frame.
For two equal-mass particles these $\hat \omega$'s are given by Eq.
(\ref{100}) below.

The relativistic relative momentum of the two particles with momentum
$\vec p$ and $-(\vec p+\vec q)$ (for the upper vertex in the homegeneous
term of Fig. 1) is given by\cite{MACH,RT}
\begin{equation}
\vec {\cal P} = \vec p +\rho (|\vec p|, |\vec q|, \theta) \vec q.
\label{3} \end{equation}
Note that $\vec {\cal P}$ is also the momentum of a particle in the
two-particle c.m. system.
Here $\theta$ is the angle between $\vec p$ and $\vec q$ and the function
$\rho$ is given by\cite{MACH,RT}
\begin{equation}
\rho (|\vec p|, |\vec q|, \theta) = \xi_q ^{-1/2}\left(\omega_{\vec p}
+\frac{\vec p.\vec q}{\xi_q ^{1/2}+\omega_{\vec p}+\omega_{\vec p +\vec q}}
\right), \label{4} \end{equation}
where $\xi_q = (\omega_{\vec p}+\omega_{\vec p +\vec q})^2 - |\vec q|^2$.
Note that in the nonrelativistic limit $\rho = 1/2$ and one has the usual
relative momentum  $\vec {\cal P} = \vec p + \vec q/2$. A completely
analogous expression exists for $\vec p$ and $\vec q$ interchanged.

In the c.m. frame of the two-particle subsystem,  where Eq. (\ref{2}) is
valid, obviously, the two particles have equal and opposite momentum $\vec
{\cal P}$ and -$\vec {\cal P}$, given by Eq. (\ref{3}),
and consequently for two equal-mass particles,
\begin{equation}
\hat \omega_{\vec p} = \hat \omega_{\vec p+\vec q}
= ({|\vec {\cal P}|} ^2 + m ^2)^{1/2}
\equiv \omega_{\vec {\cal P}}.
\label{100} \end{equation}
As a result the AT propagator is rewritten  as
\begin{equation}
G =  \frac  {2} {\omega_{\vec {\cal P}}}
\frac {2} {(\sqrt s - \omega_{\vec q})^2 - {|\vec q|}^2 - 4
 \omega_{\vec {\cal P}}^2+i0}.
\label{5} \end{equation}
This propagator is, however,  to be evaluated in the c.m. system of the
two-particle subsystem. The relevant phase space in this system
is given by $d\vec {\cal P}/
(2\pi)^3$. Consequently, the AT three-dimensional reduction of the
homogeneous part of  the equation depicted in Fig. 1, in the c.m. frame of the
two-particle  subsystem, is given by
\begin{equation}
2B\tau T = \frac {2}{(2\pi)^3}\int \frac {d\vec {\cal P}}{2\omega_{\vec
{\cal P}}} \frac {2} {(\sqrt s - \omega_{\vec q})^2 - {|\vec q|}^2 - 4
\omega_{\vec {\cal P}}^2+i0}
g\tau g T.
\label{6} \end{equation}
In Eq. (\ref{6}) the vertex function
$g$, the pair propagator $\tau$ and the $t$
matrix $T$ are to be expressed in terms of the momentum variables in the
c.m. frame of the two-particle subsystem.

Usually, the relativistic three-particle scattering equations are
conveniently written in the three-particle c.m. system. For that purpose
one has to make a transformation of variables $\vec {\cal P} \to
\vec p$ in Eq.
(\ref{6}) where $\vec {\cal P}$ and $\vec p$ are related by Eq. (\ref{3}).
As a first step of making this transformation we note that the expression
for $\rho$ given by Eq. (\ref{4}) can be conveniently rewritten in the
following two useful forms
\begin{eqnarray}
\rho (|\vec p|, |\vec q|, \theta) & = & \frac{ \sqrt \xi_q +2\omega_{\vec p}}
{2(\sqrt \xi_q +\omega_{\vec p}+\omega_{\vec p+\vec q})},  \label{7} \\
& = & \frac{1}{2}- \frac{1}{2|\vec q|^2}(\omega_{\vec p}-\omega_{\vec p
+\vec q})(\sqrt \xi_q -\omega_{\vec p}-\omega_{\vec p
+\vec q}). \label{8} \end{eqnarray}
With $\rho$ given by Eq.  (\ref{7}) the square of $\vec {\cal P}$
of Eq. (\ref{3}) is given, after some straightforward algebra, by\cite{GARC}
\begin{equation}
\vec {\cal P}^2 = \frac{1}{4}(\omega_{\vec p}+\omega_{\vec p+\vec q})^2
-\frac{1}{4}|\vec q|^2 - m^2.
\label{9} \end{equation}
Consequently, one has
\begin{equation}
4\omega_{\vec {\cal P}}^2 = (\omega_{\vec p}+\omega_{\vec p+
\vec q})^2-|\vec q|^2.
\label{20} \end{equation}

The Jacobian $J$ of the transformation $\vec {\cal P} \to \vec p$
 is given by
\begin{equation}
J \equiv det |\frac{d\vec {\cal P}}{d \vec p}|
  =1+(\vec q .\vec \nabla_{\vec p}) \rho,
\label{10} \end{equation}
 where $\vec \nabla_{\vec p}$ is the gradient with respect to $\vec p$. Using
 Eqs. (\ref{3}) and (\ref{8}) this Jacobian is evaluated after some
 straightforward algebra to yield
 \begin{equation}
  J = \frac{1}{2} \frac {\omega_{\vec p}+\omega_{\vec p+\vec q}}
  {\omega_{\vec p}\omega_{\vec p+\vec q}} \omega_{\vec {\cal P}}.
  \label{11} \end{equation}
 In  arriving at Eq. (\ref{11}) we have made use of the identity $\sqrt
 \xi_q = 2 \omega_{\vec {\cal P}}$.

Using Eqs. (\ref{20}) and (\ref{11}) , Eq. (\ref{6}) reduces to
\begin{equation}
2B\tau T = \frac{2}{(2\pi)^3} \int \frac {d\vec p}{2\omega_{\vec p}}
\frac{(\omega_{\vec p}+\omega_{\vec p
+\vec q})}{\omega_{\vec p+\vec q}[(\sqrt s-\omega_{\vec q})^2
-(\omega_{\vec p}+\omega_{\vec p
+\vec q})^2+i0]} g\tau (\sigma_{p}) g T(\vec p, \vec k; s).
\label{12} \end{equation}
Comparing Eqs. (\ref{1K}) and (\ref{12}) we identify the Born term of Eq.
(\ref{1K}) as
\begin{equation}
B(\vec q, \vec p, s) = \frac{g(\omega_{\vec p}+\omega_{\vec p
+\vec q})g}{\omega_{\vec p+\vec q}[(\sqrt s-\omega_{\vec q})^2
-(\omega_{\vec p}+\omega_{\vec p
+\vec q})^2+i0]}.
\label{1L} \end{equation}
 Note that the denominator in Eq. (\ref{12}) has two poles.
One of them given by $\sqrt s = (\omega_{\vec p}+\omega_{\vec q}+
\omega_{\vec p+\vec q})$ refers to the propagation of three particles
in the intermediate state and is responsible for maintaining the
three-particle unitarity. The other pole,  given by
$\sqrt s = (\omega_{\vec q}-\omega_{\vec p}-\omega_{\vec p+\vec q})$,
represents the propagation of a particle and two antiparticles, and does not
contribute to three-particle unitarity.

Hence the AT propagator should reduce to Eqs. (\ref{1K}) and  (\ref{1L})
in the three-particle c.m. system. The numerical applications of
Refs. \cite{TJON2,RINAT,GARC,TJON3,MACH}, however, do not use Eqs. (\ref{1K})
and (\ref{1L}) in the c.m. frame of the three-particle system. For example,
Jackson and Tjon,\cite{TJON2} Hammel et al.\cite{RINAT} Rupp and
Tjon\cite{TJON3}, and Sammarruca et al.\cite{MACH} use Eq. (\ref{6}) with
the Jacobian set equal to unity, instead of given by Eq. (\ref{11}).
Explicitly they used
\begin{equation}
2B\tau T = \frac {2}{(2\pi)^3}\int \frac {d\vec p}{2\omega_{\vec
{\cal P}}} \frac {2} {(\sqrt s - \omega_{\vec q})^2 - {|\vec q|}^2 - 4
\omega_{\vec {\cal P}}^2+i0}
g\tau (\sigma_p)g T(\vec p, \vec k; s).
\label{14} \end{equation}
in place of  Eq. (\ref{12}). Note that approximation $J=1$ is a nonrelativistic
approximation. The three particle propagator of formulation (\ref{14}) has
the correct pole, given by
$\sqrt s = (\omega_{\vec p}+\omega_{\vec q}+\omega_{\vec p+\vec q})$,
 for the propagation of three particles in the intermediate state
provided that one uses the relativistic expression (\ref{20})
for $\omega_ {\cal \vec P}$ in Eq. (\ref{14}). However,
in actual calculation the  non-relativistic approximation
$\omega_ {\cal \vec P} \simeq (m^2 + |\vec p+\vec q/2|^2)^{1/2}$ has
been used. [See, for example, Eq. (54) of Ref. \cite{TJON3} and Eq. (4) of
Ref. \cite{MACH}.] This approximation in the propagator will destroy  the
correct pole for three-particle propagation in the intermediate state,
and the resultant equation will not have the correct
relativistic threshold for scattering. Even if the proper relativistic
expression (\ref{20}) is used, Eq. (\ref{14}) will imply a wrong residue at
the pole for propagation of three particles implying violation of unitarity.
Equation (\ref{14}) is a nonrelativistic approximation
to Eqs. (\ref{1K}) and (\ref{1L}), which violates constraints
of relativistic unitarity and/or  covariance.

Garcilazo et al.\cite{GARC} used the following expression
\begin{equation}
2B\tau T = \frac{2}{(2\pi)^3} \int \frac {d\vec p}{2\omega_{\vec p}}
\frac{[(\omega_{\vec p}+\omega_{\vec p+\vec q})^2-|\vec q|^2]^{1/2}}
{\omega_{\vec p+\vec q}[(\sqrt s-\omega_{\vec q})^2
-(\omega_{\vec p}+\omega_{\vec p
+\vec q})^2+i0]} g\tau (\sigma_{p}) g T(\vec p, \vec k; s),
\label{24} \end{equation}
in place of  Eq. (\ref{12}).
This specific application of the AT propagator is distinct from Eq. (\ref{14}),
used in Refs. \cite{TJON2,TJON3,MACH}, and  uses another unnecessary
 approximation to the  Jacobian.
This approximation will have the correct pole position for the propagation of
three particles in the intermediate state but will have the wrong residue.
This will also imply  violations of conditions of relativistic unitarity.

It should be noted that the unwanted and unnecessary approximations made in
these numerical calculations
do not in any way facilitate the computational task. It is equally easy to use
the exact equations (\ref{1K}) and (\ref{1L}) in numerical applications.

In summary, we have derived the set of relativistic
three-particle equations (\ref{1K}) and (\ref{1L}) using the AT
propagator\cite{TJON1} in the
three-particle c.m. system. Originally, the AT propagator was suggested in
the c.m. frame of the two-particle subsystem. Here we provide the necessary
transformation of  the AT propagator to the three-particle c.m. system.
 We point out that the
numerical applications of this equation\cite{TJON2,RINAT,GARC,TJON3,MACH},
however, use unwanted and unnecessary nonrelativistic
approximations which do not facilitate the computational task but  violate
constraints of relativistic unitarity and/or covariance.

The work is supported in part by the Conselho Nacional de
Desenvolvimento - Cient\'\i fico e Tecnol\'ogico (CNPq) of Brasil.

\newpage
\noindent {\bf Figure Caption}

1. The diagramatic form of the three-particle scattering equation.
Each single-line represents propagation of a particle, each double-line
represents propagation of the bound state or isobar of two particles.

\unitlength=1.00mm
\special{em:linewidth 0.4pt}
\linethickness{0.4pt}
\begin{picture}(157.00,135.00)
\put(40.00,85.00){\circle{14.00}}
\put(25.00,92.00){\line(1,0){31.00}}
\put(25.00,78.00){\line(1,0){31.00}}
\put(65.00,92.00){\line(1,0){26.00}}
\put(65.00,78.00){\line(1,0){26.00}}
\put(100.00,92.00){\line(1,0){56.00}}
\put(100.00,78.00){\line(1,0){56.00}}
\put(140.00,85.00){\circle{14.00}}
\put(65.00,90.00){\line(1,0){10.00}}
\put(75.00,90.00){\line(3,-5){6.00}}
\put(81.00,80.00){\line(1,0){10.00}}
\put(100.00,90.00){\line(1,0){10.00}}
\put(110.00,90.00){\line(3,-5){6.00}}
\put(116.00,80.00){\line(1,0){19.00}}
\put(145.00,80.00){\line(1,0){11.00}}
\put(30.00,96.00){\makebox(0,0)[ct]{(P-q)}}
\put(50.00,96.00){\makebox(0,0)[ct]{k}}
\put(69.00,96.00){\makebox(0,0)[ct]{(P-q)}}
\put(86.00,96.00){\makebox(0,0)[ct]{k}}
\put(86.00,75.00){\makebox(0,0)[cb]{(P-k)}}
\put(69.00,75.00){\makebox(0,0)[cb]{q}}
\put(30.00,75.00){\makebox(0,0)[cb]{q}}
\put(50.00,75.00){\makebox(0,0)[cb]{(P-k)}}
\put(104.00,96.00){\makebox(0,0)[ct]{(P-q)}}
\put(104.00,75.00){\makebox(0,0)[cb]{q}}
\put(125.00,96.00){\makebox(0,0)[ct]{p}}
\put(125.00,75.00){\makebox(0,0)[cb]{(P-p)}}
\put(150.00,75.00){\makebox(0,0)[cb]{(P-k)}}
\put(149.00,96.00){\makebox(0,0)[ct]{k}}
\put(107.00,84.00){\makebox(0,0)[cc]{(P-p-q)}}
\put(60.00,85.00){\makebox(0,0)[cc]{$=$}}
\put(95.00,85.00){\makebox(0,0)[cc]{$+$}}
\put(45.00,80.00){\line(1,0){11.00}}
\put(25.00,90.00){\line(1,0){10.00}}
\end{picture}

\end{document}